\documentclass[multicol,twocolumn,preprintnumbers,amsmath,amssymb]{revtex4}
\usepackage{graphicx}
\usepackage{ amssymb }
\usepackage[T1]{fontenc}
\newcommand{\ket}[1]{|#1\rangle}
\newcommand{\bra}[1]{\langle#1|}
\newcommand{\abs}[1]{|#1|}

\newcommand{\ex}[1]{\langle#1\rangle}

\newcommand{\angstrom}{\textup{\AA}}

\begin{document}
\title{Blueprint of optically addressable molecular network for quantum circuit architecture}
\author{Jiawei Chang$^{1,2,3}$}\thanks{These two authors contributed equally}
\author{Tianhong Huang$^{1,2,3}$}\thanks{These two authors contributed equally}
\author{Lin Ma$^{1,2,3}$}
\author{Andrew Fisher$^4$}
\author{Nicholas M. Harrison$^5$}
\author{Taoyu Zou$^6$}\email{taoyuzou@sina.com}
\author{Hai Wang$^{1,2}$}\email{wang_lab@sina.com}
\author{Wei Wu$^{4,7}$}\email{wei.wu@ucl.ac.uk}
\affiliation{$^1$Yunnan Key Laboratory of Metal-Organic Molecular Materials and Devices Kunming University Kunming Yunnan Province P. R. China}
\affiliation{$^2$Key Laboratory of Yunnan Provincial Higher Education Institutions for Organic Optoelectronic Materials and Devices, Kunming University, Kunming, Yunnan Province, P. R. China} 
\affiliation{$^3$School of Physical Science and Technology, Kunming University, Kunming, Yunnan Province, P. R. China}
\affiliation{$^4$UCL Department of Physics and Astronomy and London Centre for Nanotechnology, University College London, Gower Street, London WC1E 6BT}
\affiliation{$^5$Department of Chemistry, Imperial College London, White City Campus, 80 Wood Lane, London, W12 0BZ, United Kingdom}
\affiliation{$^6$Yunnan Ocean Organic Optoelectronic Technology Ltd, Kunming, 650214, Yunnan Province, P. R. China}
\affiliation{$^7$UCL Institute for Materials Discovery, University College London, Gower Street, London WC1E 6BT, United Kingdom}
\date{\today}%

\begin{abstract}
Optically connecting quantum bits can effectively reduce decoherence and facilitate long-distance communication. Optically addressable spin-bearing molecules have been demonstrated to have a good potential for quantum computing. In this report optically induced exchange interactions and spin dynamics, which are inherently important for spin-based quantum computing, have been calculated for a bi-radical - a potential quantum computing circuit unit. Consistent with the previous experimental observation of spin coherence induced by optical excitation, our work demonstrated an optically driven quantum gate operation scheme, implying a great potential of molecular quantum-circuit network. A blueprint of quantum circuit, integrating two-dimensional molecular network and programmable nano-photonics, both of which have been under extensive investigations and rather mature, was proposed. We thus envisage computational exploration of chemical database to identify suitable candidates for molecular spin quantum bit and coupler, which could be optimally integrated with nano-photonic devices to realize quantum circuit. The work presented here would therefore open up a new direction to explore 'Click Chemistry' for quantum technology.
\end{abstract}


\maketitle

\section{Introduction} 



Quantum computer needs a good \textit{control} on the interaction among \textit{well-isolated} quantum bit (qubit), which should be addressed properly by optimal design \cite{nielsenchuang}. In the 'quantum computing stack' (QCS) \cite{alexeev2021}, the quality of qubit is the key foundation for control engineering, both of which form an indispensable material basis for quantum software and algorithms in the upper levels of QCS. Recent developments in QC based on superconducting circuits have demonstrated superior QC capacity in dealing with problems that are inconvenient for the classical counterparts \cite{thew2020, arute2019, wu2021, ebadi2021, huang2019, xue2021}. On the other hand, more accessible working environment and further elevation of the number of the integrated qubits with quantum error corrections \cite{alexeev2021} are important challenges yet to be conquered. Moreover, developing energy-efficient and sustainable quantum technology is crucial \cite{auf2022}, for which high-temperature (or even room-temperature) QC could be useful. Recent proposals of molecular QC platforms, which explore molecular spin qubit and optically accessible triplet state, could shed light on the concerns aforementioned \cite{bayliss2020,wasielewski2020,ma2022,wu2023,delaney2022}. 
 
\begin{figure*}[htbp]
\includegraphics[scale=0.5,clip=true, trim= 0mm 0mm 0mm 0mm]{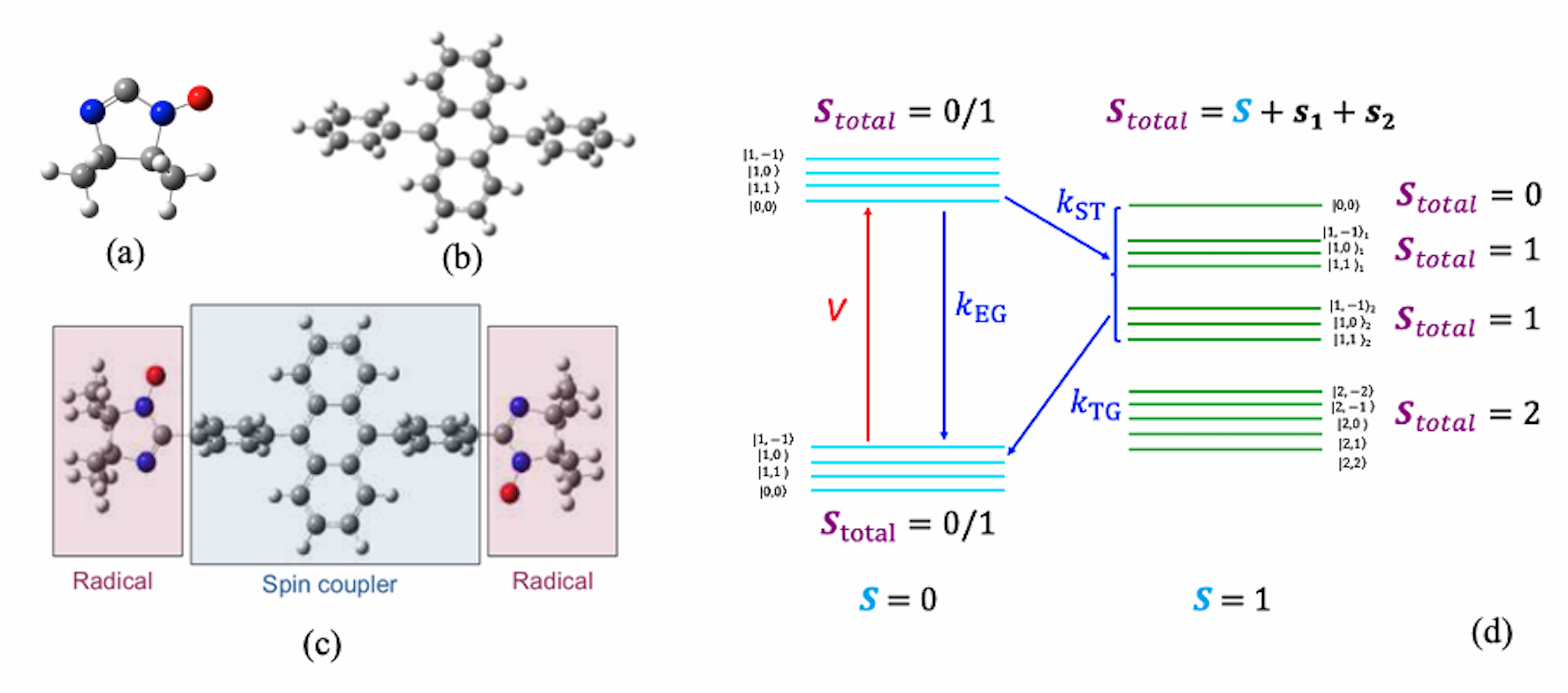}
\caption{(Color online.) One example for the MRTS systems and the spin coupling mechanism are shown. (a) TYY radical, (b) DPA spin coupler, and (c) combined double-radical triplet system. In (d), we have shown the coherent and incoherent processes. The coherent processes include spin-spin interaction among radicals and triplet and optical driving field ($V$), while the incoherent one the inter-system crossing and the decays of the triplet and singlet excited states. The decay rates $k_{ST}$, $k_{TG}$, and $k_{EG}$ are also illustrated.}\label{pic:dpatyy}
\end{figure*}

Organic radical, containing $\frac{1}{2}$-spins, is a natural realisation of qubit  \cite{rv2012, ji2020}. An example for the organic radical, 4,4,5,5-tetramethyl-1-yloxyimidazolin-2-yl (TYY), is shown in Fig.~\ref{pic:dpatyy}a. The spin relaxation and coherence times of organic radicals can be very long (up to seconds) even at high or room temperature due to weak spin-orbit coupling and hyperfine interaction \cite{warner2013,ga2019, atzori2016, atzori2019, Kjaergaard2020}. For multi-qubit gate operations, a promising route to control the interaction between radicals is optical excitation. A molecule consisting of two or more radicals linked by the optically active spin coupler such as bi-radical could be an ideal platform for demonstration of optically controlled quantum gate operation. For example, di-phenyl anthracene (DPA) molecule, shown in Fig.\ref{pic:dpatyy}b, can be excited from a singlet ground state to a triplet state due to the inter-system crossing (ISC)~\cite{porphyrinbooks}, thus mediating the interaction among radicals. The spin coupling mechanism is shown in Fig.\ref{pic:dpatyy}d. A multi-radical-triplet system (MRTS) contain radicals linked by spin coupler, such as the biTYY-DPA \cite{teki2000}, as shown in Fig.\ref{pic:dpatyy}c. 'Click chemistry', which has been extremely successful in biochemistry and drug discovery, could provide us with a superior methodology to assemble optimal radicals and couplers for quantum circuit \cite{kolb2003}. 

Extensive experimental efforts, combining time-resolved electron paramagnetic resonance (TREPR) and optical excitations~\cite{ishii1998,corvaja2000,teki2000, franco2006, sato2007,zhang2022} have been made to control exchange interactions in MRTS, such as biTYY-DPA (Fig.\ref{pic:dpatyy}c). This spin coupling mechanism offers a promising route to control quantum gate operations at a time scale of $\sim ns$. Most of the TREPR experiments can be performed at the high temperature (77 K, the boiling point of liquid nitrogen), which could facilitate the development of more practical quantum technologies. Previously, a first-principles calculation was carried out for the spin-$2$ state of the biTYY-DPA molecule and a qualitative model for the spin dynamics in MRTS has been proposed~\cite{huai2005}. To the authors' best knowledge, there have been few reports about (i) quantitative calculation of the exchange interaction between the transient triplet and radicals in MRTS and (ii) the simulation of the TREPR spectra using the theory of open quantum systems \cite{bp2002}, both of which are the focus of this work. Our theoretical findings are in a good agreement with the previous experiment. Moreover, we have further proposed a molecular architecture integrated with nanophotonic devices to scale up the quantum circuits, thus paving the way towards optically driven 'Click Chemistry' molecular circuit.



\section{Computational details}\label{sec:methods}
\subsection{First-principles density-functional-theory calculations for exchange interactions}
First-principles calculations have been carried out using hybrid exchange density function theory (HDFT) with a 6--31G basis set  in the Gaussian 09 code \cite{gaussian09}. The self-consistent field (SCF) procedure was converged to a tolerance of $10^{-6}$ a.u. ($\sim 0.3$ Kelvin). The broken-symmetry method \cite{wu2013} was used to allow spins to localize on the radicals and the convergence to a low-spin configuration. Electronic exchange and correlation are described using the B3LYP hybrid exchange density functional \cite{b3lyp}, the advantages of which include a partial elimination of the electronic self-interaction error and a balance of the tendencies to delocalize one-electron wave-functions. The B3LYP density functional has previously been shown to provide an accurate description of the electronic structure and magnetic properties for both inorganic and organic compounds \cite{wu2015, zou2018}. 

Supposing that an MRTS consists of two radicals and one spin coupler,  and that the exchange interaction dominates over both dipolar interactions and spin-orbit coupling, then the Heisenberg spin Hamiltonian before optical excitations reads
\begin{equation}\label{eq:rrheisenberg}
\hat{H}_0=J_0\hat{\vec{s}}_1\cdot\hat{\vec{s}}_2,
\end{equation} where $J_0$ is the exchange interaction between radical spins $\hat{\vec{s}}_1$ and $\hat{\vec{s}}_2$, which can be computed as the energy difference between the triplet and broken-symmetry (BS) states of radicals, $J_0=2(E_{\mathrm{triplet}}-E_{\mathrm{BS}})$.
When the spin coupler is in a triplet state after optical excitations, the spin Hamiltonian changes to
\begin{equation}\label{eq:rtpheisenberg}
\hat{H}_1=J_1\hat{\vec{s}}_1\cdot\hat{\vec{S}}+J_2\hat{\vec{S}}\cdot\hat{\vec{s}}_2+J_3\hat{\vec{s}}_1\cdot\hat{\vec{s}}_2,
\end{equation} where $J_1$ ($J_2$) is defined as the exchange interaction between radical spins $\hat{\vec{s}}_{1}$ ($\hat{\vec{s}}_{2}$) and the triplet spin $\hat{\vec{S}}$. $J_3$ is the exchange interaction between two radical spins after optical excitations. To compute $J_1$, $J_2$, and $J_3$, we first find the total energies of the states $\ket{a}=\ket{\uparrow\Uparrow\uparrow}$, $\ket{b}=\ket{\uparrow\Uparrow\downarrow}$, $\ket{c}=\ket{\downarrow\Uparrow\downarrow}$, and $\ket{d}=\ket{\downarrow\Uparrow\uparrow}$, where $\uparrow$($\downarrow$) is defined as the spin state $\ket{s=\frac{1}{2},s_z=\frac{1}{2}}$ ($\ket{s=\frac{1}{2},s_z=-\frac{1}{2}}$) of a radical and $\Uparrow$ as $\ket{S=1,S_z=1}$ for the spin-coupler triplet.  Then we can have
\begin{eqnarray}\label{eq:exchange1}
J_1&=&(\Delta E_{ac}+\Delta E_{bd})/2,\nonumber
\\J_2&=&(\Delta E_{ac}-\Delta E_{bd})/2,\nonumber
\\J_3&=&\Delta E_{cd}+\Delta E_{ab},
\end{eqnarray} 
where $\Delta E_{ij}=E_i-E_j$ is defined as the energy difference between the states with the spin configurations $i$ and $j$ ($\in$ \{$a, b, c, d$\}). 

Note that the broken-symmetry singlet and triplet states formed by two radical spins alone have the same $z$-component of spin as $\ket{c}$ and $\ket{b},\ket{d}$, respectively, but much lower energies, owing to the triplet excitation energy ($\sim$ eV in general). The Kohn-Sham orbitals therefore need to be repopulated in order to drive the system into $\ket{b}$, $\ket{c}$, or $\ket{d}$. If two radicals are symmetric, then $E_b=E_d$ and therefore
\begin{eqnarray}\label{eq:exchange2}
J_1&=&J_2=\Delta E_{ac}/2,\nonumber
\\J_3&=&\Delta E_{cb}+\Delta E_{ab}.
\end{eqnarray}
The expectation value of the total spin operator $\hat{\vec{S}}_{\mathrm{total}}=\hat{\vec{s}}_1 +\hat{\vec{S}}+\hat{\vec{s}}_2$ is $6$ for the state $\ket{a}$ (since it is a total spin eigenstate with $S_\mathrm{total}=2$), 3 for $\ket{b}$ and $\ket{d}$, and 2 for $\ket{c}$. We adopted that the exchange interactions are antiferromagnetic (AFM) when $J > 0$ while ferromagnetic (FM) when $J < 0$. The spin Hamiltonians in eq.\ref{eq:rrheisenberg} and eq.\ref{eq:rtpheisenberg} and the computation of the exchange interactions therein can be easily generalized to the molecular structure with more radicals.


Here $J_1$ is equal to $J_2$ because the biTYY-DPA molecule has an inversion symmetry. $J_0$, $J_1$ and $J_3$ were computed for the dihedral angles between the phenyl ring and the anthracene, ranging from $0^\circ$ (coplanar) to $90^\circ$ (perpendicular) with $10^\circ$ increments. In all the calculations, we freezed the TYY radicals and rotate only the phenyl ring, which meant we assumed the dihedral angles at each end of the coupler are equal. 

\subsection{The time evolution of the density matrix and TR-EPR spectra simulations}\label{sec:treprmethod}

\subsubsection{System Hamiltonian}

After taking into account the optical driving field, our total Hamiltonian reads
\begin{widetext}
\begin{eqnarray}\label{eq:system}
\hat{H}_{opt} &=& \ket{S_0}[\hat{H}_0+g_r\mu_B\vec{B}\cdot(\vec{s}_1+\vec{s}_2)]\bra{S_0} +\nonumber
\\&&\ket{S_1}[\hat{H}_0+g_r\mu_B\vec{B}\cdot(\vec{s}_1+\vec{s}_2)]\bra{S_1}+\nonumber
\\&&\ket{T_1}[\hat{H}_1+g_r\mu_B\vec{B}\cdot(\vec{s}_1+\vec{s}_2)+g_c\mu_B\vec{B}\cdot \vec{S}+D\hat{S}_z^2+E(\hat{S}_x^2-\hat{S}_y^2)]\bra{T_1}+\nonumber
\\&&V(\ket{S_0}\bra{S_1}+\ket{S_1}\bra{S_0}).
\end{eqnarray}
\end{widetext}

Here $S_0$ is the singlet ground state, $S_1$ the first singlet excited state, and $T_1$ the triplet ground state. $D$ and $E$ are the zero-field splittings (ZFS) for the triplet. $g_r$ is the g-factor for the radical spin, while $g_c$ for the coupler spin. Here $V$ is the transition matrix element between the ground and excited singlet states ($S_0$ and $S_1$) due to the optical driving field. $\vec{B}$ is a static magnetic field and $\mu_B$ is the Bohr magneton. $\hat{H}_0$ and $\hat{H}_1$ are defined in eq.\ref{eq:rrheisenberg} and \ref{eq:rtpheisenberg}, respectively. Here we assume that the exchange interaction between radicals $J_0$ is unchanged in the singlet excited states.

\subsubsection{Quantum jump operators due to the system-environment couplings}
To include the environment effects on the time evolution of the density matrix, we used the super operators (the Liuvillian) within the Markovian approximation, leading to the Lindblad formalism. These super operators read
\begin{equation}\label{eq:quantumjumps}
\hat{\hat{L}}_i\hat{\rho}=\sum_{\mu=1}^{n_i}\gamma^{\mu}_i[\hat{l}^\mu_i\hat{\rho}\hat{l}^{\mu\dagger}_{i}-\frac{1}{2}\{{\hat{\rho},\hat{l}^{\mu\dagger}_{i}\hat{l}^\mu_{i}}\}]
\end{equation}

Here $\hat{\rho}$ is the density matrix for MRTS. The incoherent processes are described by $\gamma^{\mu}_i$ (the decoherence rate) and $\hat{l}^{\mu}_i$ (the quantum jump operator), where $i$ labels different decoherence processes and $\mu$ the operators in the process. And $n_i$ is the number of the operators in each physical process $i$. For our system, $i$ = 1 to 5. $\hat{\hat{L}}_{1,2}$ with $n_{1,2}$ = 3 is used to describe the relaxation of the radical spins $\vec{s}_1$ and $\vec{s}_2$ as follows, $\hat{l}_1^{1,2}$ = $\hat{s}^-_{1,2}$ (spin-$\frac{1}{2}$ lowering operator), $\hat{l}_2^{1,2}$ = $\hat{s}^+_{1,2}$ (spin-$\frac{1}{2}$ raising operator), and $\hat{l}_3^{1,2}$ = $\hat{s}^z_{1,2}$(the $z$-component of the spin-$\frac{1}{2}$). The $\hat{s}^z$ operator is responsible for the pure dephasing of quantum states, whereas $\hat{s}^+$ and $\hat{s}^-$ for quantum jumps. $\hat{\hat{L}}_2$ with $n_2$ =8 is used as a super operator to describe the relaxation of the triplet state. We have used the corresponding eight Gell-Mann matrices for these operators. Moreover, we have taken into account the spin-spin (transverse spin) relaxation \cite{ivady2020} within the mean-field approximation induced by the interaction between the spins in the system and those in the environment, i.e., $\hat{s}_1\hat{s}_2\sim \ex{s}_1\hat{s}_2$. Hence the $T_2$ relaxation processes can be absorbed into the Lindblad formalisms $\hat{\hat{L}}_{1,2}$ described above.

Here, we have taken into account the ISC in $\hat{\hat{L}}_3$, assuming that the total spin angular momentum and its $z$-component are both conserved, i.e., $S_\mathrm{total} = 0$ or $1$ (twice), thus obtaining the following operators. $\hat{\hat{L}}_3$ with $n_3$ = 7 describes the transitions from the single excited state to the triplet state. For example, some of the operators read $\hat{l}_3^{1}=\ket{0,0}_{T_1}\bra{0,0}_{S_1}$ and $\hat{l}_3^{2}=\ket{1,-1}_{T}^1\bra{1,-1}_{S_1}$. Here we use $\ket{S_{total}, S_{total,z}}$ to represent our states formed in the triplet ($T_1$) and the $S_1$ manifolds as shown in Fig.\ref{pic:dpatyy}(d). Similarly we can define the decay operator $\hat{\hat{L}}_4$ ($n_4=7$) for the transition from the triplet state to the singlet ground state. For example, some of the operators read $\hat{l}_4^{1}=\ket{0,0}_{S_0}\bra{0,0}_{T}$ and $\hat{l}_4^{2}=\ket{1,-1}_{S_0}\bra{1,-1}_{T}$. The fifth super-operator $\hat{\hat{L}}_5$ is responsible for the spontaneous decay of the singlet excited state ($S_1$) down to the ground state ($S_0$). There are four operators ($n_5=4$), including $\hat{l}_5^{1}=\ket{0,0}_{S_0}\bra{0,0}_{S_1}$, $\hat{l}_5^{2}=\ket{1,-1}_{S_0}\bra{1,-1}_{S_1}$,$\hat{l}_5^{3}=\ket{1,0}_{S_0}\bra{1,0}_{S_1}$, and $\hat{l}_5^{4}=\ket{1,1}_{S_0}\bra{1,1}_{S_1}$. All the coupling parameters $\gamma^{\mu}_i$ ($i$ from 1 to 5) have been simplified by using a single parameter $\gamma_{\mathrm{radical}}$, $\gamma_{\mathrm{triplet}}$, $k_{st}$, $k_{tg}$, and $k_{eg}$ for these five incoherent processes. The lifetime of the triplet, characterized by the parameter $k_{tg}$, is usually long ($ms$) because the decay to the singlet ground state is forbidden due to spin momentum conservation and hence dominated by non-radiative processes.

\subsubsection{The total Liuvillian and the simulation of TREPR spectra}
Therefore the total Liuvillian operator can be written as follows.

\begin{equation}\label{eq:total}
\frac{d\hat{\rho}}{dt}=\hat{\hat{\mathcal{L}}}\hat{\rho}=-i[\hat{H}_{opt},\hat{\rho}]+[\sum_{i=1}^5\hat{\hat{L}}_i]\hat{\rho}
\end{equation}

Here the first part is the coherent interaction from the effective spin Hamiltonian and optical field (see eq.\ref{eq:system}), which is the commutator between the Hamiltonian and the density matrix. The second part includes the incoherent processes associated with relaxations and crossovers between states (see eq.\ref{eq:quantumjumps}). Therefore, we have 20 by 20 Hamiltonian matrix (4 for the singlet ground and excited states with radicals, 12 for the triplet ground states with radicals), leading to a 400 by 400 Liuvillian.

The TR-EPR spectra can be computed as follows \cite{blankbook,misrabook},

\begin{eqnarray}
&&I_0(t,\omega,\theta,\phi)=\\\nonumber
&&\abs{\mathrm{Tr}\{\hat{\rho(t)}S_{x}^{\mathrm{total}}(\theta,\phi)[i\hat{\hat{\mathcal{L}}}(\theta, \phi)-\omega I]^{-1}S_{x}^{\mathrm{total}}(\theta, \phi)\}}.
\end{eqnarray}
Here $S_{\mathrm{mw}}^{\mathrm{total}} =\vec{S}^{\mathrm{total}}\cdot{\vec{e}_{\mathrm{mw}}}$, where $\vec{e}_{\mathrm{mw}} = (-\sin{\phi},\cos{\phi},0)$, is the total spin component along the microwave field direction, which is alway perpendicular to the static magnetic field along $\vec{e}_{\mathrm{static}}=(\sin{\theta}\cos{\phi},\sin{\theta}\sin{\phi},\cos{\theta})$ direction, and $\omega$ is the microwave field frequency.

We have also computed the powder-averaged TREPR spectra as follows,

\begin{equation}
I(t,\omega) = \int_{0}^{\pi}d\theta\int_{0}^{2\pi}d\phi I_0(t,\omega,\theta,\phi).
\end{equation}
Numerically we have chosen 50 points uniformly along $\theta$ and 100 points uniformly along $\phi$ (5000 points in total) as done in Easyspin \cite{easyspin}. 

\section{Results and discussion}
\subsection{DFT calculations for exchange interactions}\label{subsec:dft}
\begin{figure*}[htbp]
\includegraphics[scale=0.5,clip=true, trim= 0mm 0mm 0mm 0mm]{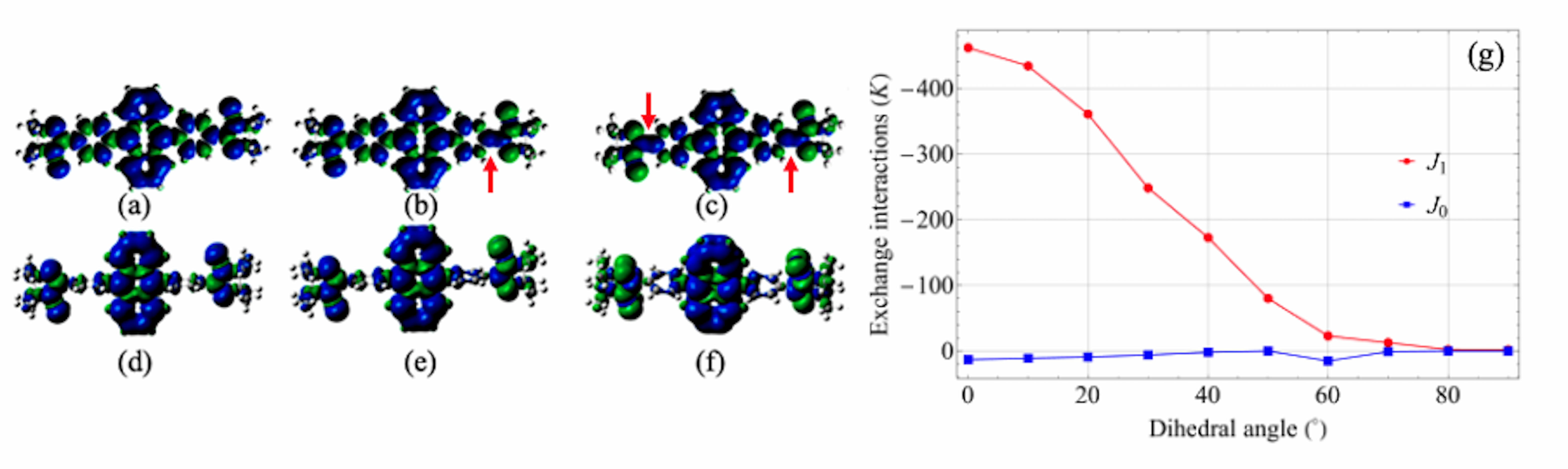}
\caption{(Color online.) The isosurfaces of spin densities in biTYY-DPA with the dihedral angle of $0^\circ$ ($90^\circ$) with spin configurations $\ket{a}$, $\ket{b}$, and $\ket{c}$ are shown in (a) - (c) ((d)-(f)). Positive is in blue and negative in green. The isosurface value is set to $0.01 e/\mathrm{\angstrom}^3$. In (g), the exchange interactions $J_0$ (blue squares) and $J_1$ (red circles) are plotted as functions of the dihedral angle between the anthracene and the phenyl ring. Notice that for all the angles studied here, the sign of the exchange interaction is switched by optical excitation and ISC, and that $J_1$ decreases as the dihedral angle increases owing to the suppression of the triplet wave function on DPA. On the other hand, the dependence of $J_0$ on the dihedral angle is more complicated because the exchange interaction between radicals is through spin polarisation.}\label{pic:dft}
\end{figure*}

When the spin coupler is in its ground state, the two TYY radicals interact weakly through the spin polarisation (induced by the radical spin) on the coupler. After the optical excitation and ISC, the spin polarisation on the coupler is mostly dominated by the triplet spin, as shown in Fig.\ref{pic:dft}(a-f). In our calculations, $J_3$ is three to four orders smaller than $J_0$ for all the dihedral angles studied (Fig.\ref{pic:dft}g). We therefore neglect $J_3$ in what follows. When the dihedral angle is $0^\circ$, $J_0/k_B$ is predicted to be $16.8$ K and  $J_1/k_B$ to be $-461.4$ K, which is above the room temperature. When the dihedral angle is $60^\circ$, which is a geometry that has been observed in the DPA crystal structure~\cite{dpaangle}, $J_0/k_B$ is predicted to be $15.2$ K and $J_1/k_B$ to be $-22.8$ K. When the dihedral angle is $90^\circ$, $J_0$ is computed to be negligible, however the dipolar interaction between the spins is estimated to be $\sim 10^{-4}$ K as the distance between radicals is $\sim 17 \  \angstrom$, and $J_1/k_B$ is predicted to be $-2.0$ K. The calculation results are consistent with the TREPR experiments for the optically driven transient magnetic properties of biTYY-DPA, which confirms the radical spins are aligned after the optical excitation as shown in the previous experiment~\cite{teki2000}. At suitable temperatures the optical excitation and ISC can change the spin alignment of the radicals from AFM to FM with enhanced magnitude. This implies that biTYY-DPA not only has potential for quantum gate operations but also can work as an optically controlled single-molecule magnetic switch. This can potentially work at room temperature if the molecule can be prepared with a  dihedral angle smaller than $20^\circ$ on a surface, as suggested in Fig.\ref{pic:dft}g. The calculated spin densities reveal the nature of the exchange interactions between the radicals and the triplet on DPA. The spin densities of the three spin configurations $\ket{a}$, $\ket{b}$, and $\ket{c}$ for the dihedral angle of $0^\circ$ ($90^\circ$) are shown in Fig.~\ref{pic:dft}a-c (d-f). $J_1$ decreases as the dihedral angle is increased because the rotation of the phenyl ring away from the coplanar geometry suppresses the delocalisation of the $\pi-$orbitals of DPA, which dominate the wave function of the triplet exciton. This is manifested by the much smaller spin densities on phenyl rings with the dihedral angle equal to $90^\circ$ than $0^\circ$, as shown in Fig.~\ref{pic:dft}(d-f). The effect of phenyl rings on spin densities implies that we could engineer molecules to control spin densities thus interaction strength. In contrast to these large variations on the phenyl rings the spin densities on the radicals are insensitive to the dihedral angle: Mulliken population analysis yields a spin moment of $\sim 0.3 \mu_B$ on the nitrogen and $\sim 0.5 \mu_B$ on the oxygen for all the dihedral angles, which indicates the radical spins are well preserved although we have optical excitation on the spin coupler.

The mechanism for the strong radical-triplet exchange interaction $J_1$ can be understood using the Ovchinnikov's topological spin alternation rule for $\pi$-conjugated systems~\cite{ovchinnikov1978, lieb1989}: the spin-up and spin-down densities alternate on the neighbouring carbon atoms. The underlying physics of this rule is related to the formation of the spin polarisation induced by electron delocalisation in a $\pi$-conjugated system, similar to the indirect exchange mechanism in inorganic insulators~\cite{anderson} and the Lieb's theorem for the bipartite graphene lattice~\cite{lieb1989}. In the $\ket{a}$ state, the spin densities are consistent with the Ovchinnikov's topological spin alternation rule, whereas in the $\ket{b}$ and $\ket{c}$ states the spin alternation pattern is violated when the spin-up densities meet at the junction between the phenyl ring and the TYY radicals (highlighted by the red arrows in Fig.\ref{pic:dft}b,c). Therefore, the spin-aligned state $\ket{a}$ is favoured and the exchange interaction for the triplet excited state on DPA is FM. This mechanism based on the Ovchinnikov's rule could be used to engineer the sign of the exchange interaction.


\subsection{Spin dynamics with optical excitations and TREPR spectra simulations}
\begin{figure*}[htbp]
\includegraphics[scale=0.7,clip=true, trim= 0mm 0mm 0mm 0mm]{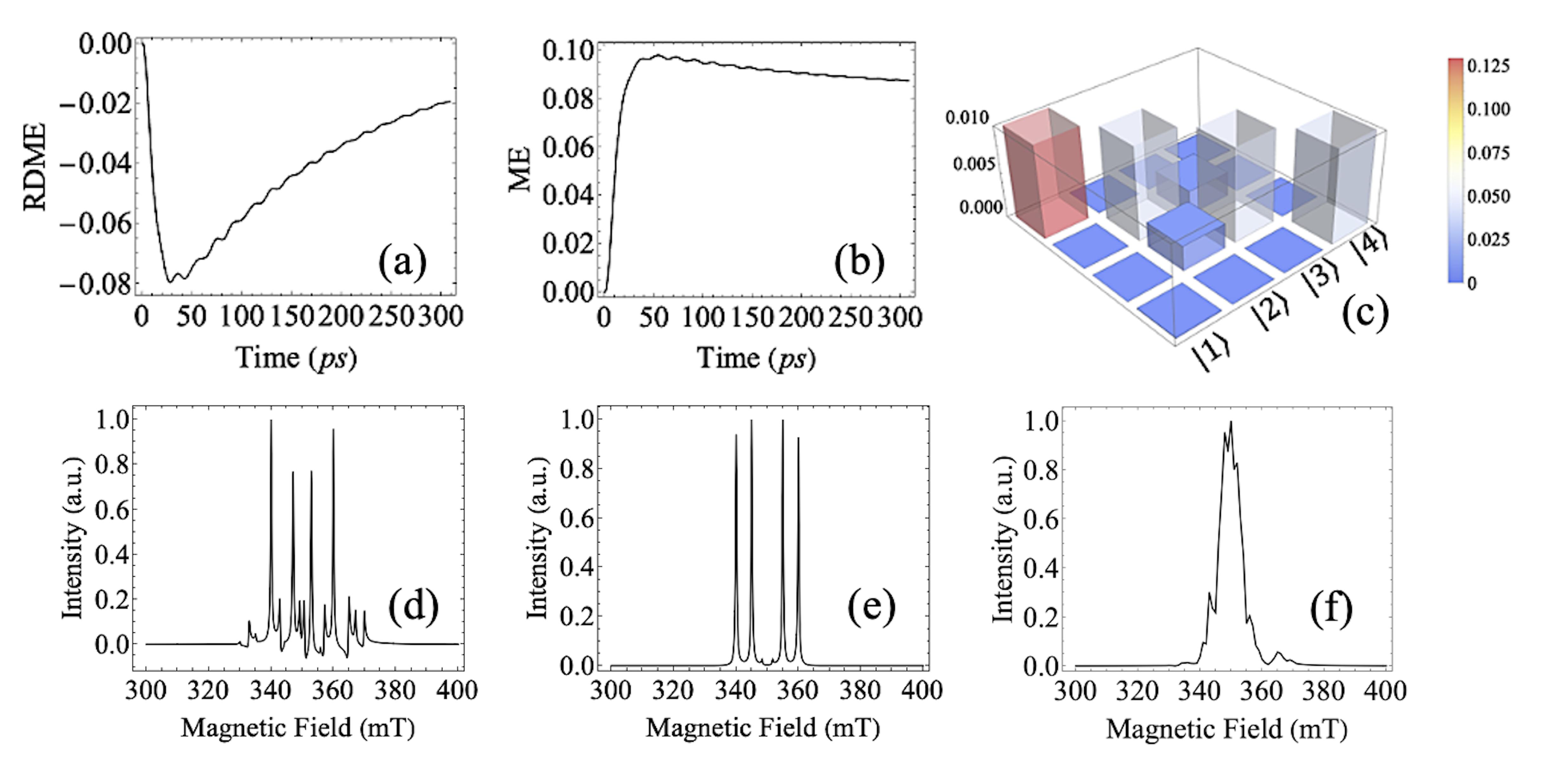}
\caption{(Color online.) The time evolutions of the density matrix elements are shown. In (a), we show the off-diagonal term ($\ket{\frac{1}{2},\frac{1}{2}}\bra{\frac{1}{2},-\frac{1}{2}}$) of the reduced density matrix element (RDME) for two radicals, which indicates the entanglement between two radicals created during this process. In (b), we show the populations for the state $\ket{\frac{1}{2},1,-\frac{1}{2}}$, which has a Rabi-oscillation behaviour. In (c), we have computed the magnitudes of the reduced density matrix for two radicals as a tomography. Notice that we only plotted up to 0.01. We show the EPR spectra at early stage with different exchange interactions between radical and triplet: (d) $J_1=-10$ mT , (e) $-10^3$ mT, and (f) $-10^5$ mT.}
\label{pic:densitymatrix}
\end{figure*}

We have also computed the dynamics of the whole system based on the theory of open quantum systems described in \S\ref{sec:treprmethod}. In Fig.\ref{pic:densitymatrix} (a), we show the time evolution of the off-diagonal term (coherence) for the reduced density matrix of the two radicals ($\ket{\frac{1}{2},\frac{1}{2}}\bra{\frac{1}{2},-\frac{1}{2}}$), after tracing out the triplet manifold. We can see that the magnitude of the matrix element will increase when applying the optical driving field, and then will decay slowly as the triplet vanishes. This suggests the exchange interaction between radical and triplet induced will trigger the coherence or entanglement between the two radical spins. The diagonal term (population) for the $\ket{\frac{1}{2},1,-\frac{1}{2}}$ state in the density matrix (Fig.\ref{pic:densitymatrix}b) shows a decaying Rabi-oscillation characteristics, indicating the coherence created during this process. In Fig.\ref{pic:densitymatrix} (c), the tomography for the magnitudes of the reduced density-matrix of the two radicals at the early stage ($T = 6.2 \ ps$) of the time evolution after applying the optical excitation has been plotted, in which we can clearly see the nonzero off-diagonal term for spin coherence. Here $\ket{1} = \ket{\frac{1}{2},\frac{1}{2}}_L\ket{\frac{1}{2},\frac{1}{2}}_R$, $\ket{2} = \ket{\frac{1}{2},\frac{1}{2}}_L\ket{\frac{1}{2},-\frac{1}{2}}_R$, $\ket{3} = \ket{\frac{1}{2},-\frac{1}{2}}_L\ket{\frac{1}{2},\frac{1}{2}}_R$, and $\ket{4} = \ket{\frac{1}{2},-\frac{1}{2}}_L\ket{\frac{1}{2},-\frac{1}{2}}_R$ ($L$ and $R$ label the radical on the left and right, respectively). We have also investigated the TREPR spectra at the early stage ($T = 6.2 \ ps$) in Fig.\ref{pic:densitymatrix}(d-f). We have used different parameters for the exchange interaction between radical and triplet, which are comparable to the range of the exchange interactions computed by DFT as shown in \S\ref{subsec:dft}. The exchange interaction between triplet and radical is $-10$ and $-10^3$ mT for Fig.\ref{pic:densitymatrix}(d) and (e), respectively. We have also computed the powder-averaged spectra as shown in Fig.\ref{pic:densitymatrix}(f). The main and side peaks have successfully been reproduced based on our open-quantum-system modelling, as compared with the experimental results \cite{teki2000}. Notice here that we haven't distinguished emission and absorption. For the static magnetic field (X-band), the spin anisotropy, spin relaxation rates, the ISC rate, and spontaneous decay rate, we have used the same parameters as the previous work \cite{ma2022} When the exchange interaction is smaller than the static magnetic field ($\sim350$ mT), there are more features in the spectra than those from the much larger exchange interaction. The side peaks stem from the triplet state of DPA while the central peaks (when the exchange interaction is -10 mT) originate from the radical spins, which are renormalized by the interactions with the triplet on DPA. Our EPR simulations are qualitatively also consistent with the experimental results in Ref.\cite{teki2000}, except that the authors therein distinguished the absorption and emission. The EPR spectra for the larger exchange interactions are saturated due to the limitation of the magnitude of the static magnetic field, which implies the EPR spectra will be dominated by the exchange interaction between the radicals and the triplet. Therefore, in this range of static magnetic field, we can only see the effect from the spin anisotropy ($D$ and $E$). For the large exchange interaction, we need other instruments to probe them, such as neutron scattering \cite{ns}.

\subsection{QCAD based on molecules}
\begin{figure*}[htbp]
\includegraphics[scale=0.65,clip=true, trim= 0mm 5mm 0mm 0mm]{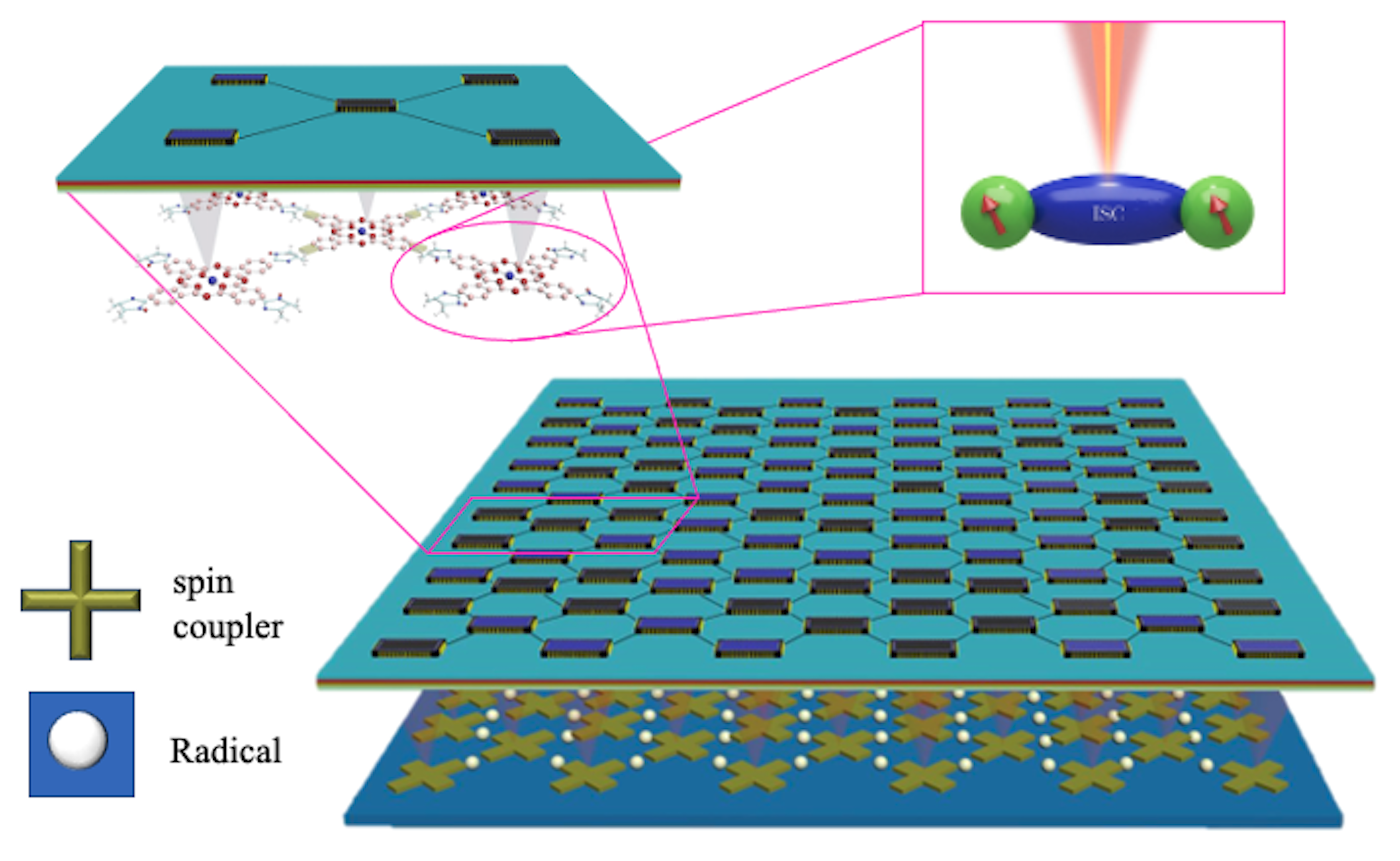}
\caption{(Color online.) A QC architecture based on molecules and optical instruments is shown as an illustration for a more complex molecular QC network. The fundamental physical mechanism is through the optical excitation and ISC, coupling radicals, as computed in the previous sections. Here radicals (TYY as an example) are linked by the spin couplers (phthalocyanine molecules as an example), which could be either achieved by Van der Waals forces or through chemical bonds by designing bridging molecular structures. And the optical devices working as the control unit need to be integrated with the molecular spin couplers, such that we can program them as needed to control a quantum gate operation \cite{arute2019}.}
\label{pic:structures}
\end{figure*}

Quantum computer-aided design (QCAD) has been proposed recently for the quantum computing architectures based on silicon quantum dots \cite{gao2013, niquet2020, kyaw2021}, which has a huge potential to optimise the quantum circuit design, thus improving the design efficiency, although it is still in the early stage. In the mean time, the research on two-dimensional (2D) materials have been surging rapidly \cite{gibertini2019,ferrari2015, zhuang2015}. We could therefore generalize bi-radical to an example for molecular quantum-circuit architecture as shown in Fig.\ref{pic:structures}. A two-dimensional molecular network consists of TYY radicals (qubits) and phthalocyanine (Pc, working as a spin coupler), which can be either chemically bonded by suitable bridging molecular structures or linked through Van der Waals forces. Here we have chose Pc as a spin coupler because (i) it has good optical properties \cite{porphyrinbooks} and (ii) as a planar molecule, it is great building block to form two-dimensional network. On top of this molecular network, we could build nanophotonic devices controlling the spin and electronic states of the Pc molecules, i.e. inducing a transition from the singlet to the triplet, thus mediating the interactions among the radicals. The fabrication of the optical devices has been demonstrated recently by using an optical twizzer \cite{dsp2021}. Moreover, we can explore the triplet excited states, i.e., promoting the triplet from the ground state to the excited state, which is expected to allow us to further separate the radicals, ideally tens of nanometer distances, thus easing the design and fabrication of optical devices. We could not only take advantages from the recent development of the programable photonic quantum circuits \cite{bogaerts2020}, but also integrate spin qubits with photonics \cite{elshaari2020}. Regarding the feasibility of fabricating optical devices, one choice is to use nanometer-size single photon emitter recently developed such as the gold tips or nanoscale gaps on WSe$_2$ \cite{kern2016, ziegler2018, peng2020}. In addition, the quantum calligraphy, in which encoding strains into the 2D materials can be used to create and locate the single-photon emitters with a precision of nanometers \cite{rosenberger2019}. Also, the nanometer-sized single photon emitters can be fabricated by using quantum dots \cite{liu2021}.  Another alternative for the optical devices is to use organic molecules such as Pc and dibenzoterrylene that have very good optical emission properties \cite{porphyrinbooks, toninelli2021}, which is also compatible with spin couplers, thus further easing the integration with the molecular network. As shown in the above sections, the mediating couplings and quantum dynamics have been calculated, which suggests this proposal is theoretically feasible. The persistence of the exchange interaction between the radical and the triplet could be problematic for read-out. Further work will therefore be required either to identify mechanisms that can turn off the interaction after a defined time in order to read out quantum information, or to understand how to exploit the always-on interaction to perform quantum computation \cite{benjamin2003}. 




\section{Conclusions}
In this work, we have computed the exchange interaction between the radicals and optically induced triplets in biTYY-DPA from first principles, which are consistent with the previous experimental results observing the spin alignment with optical excitations. We find the radical-triplet exchange interactions $J_1$ are enhanced very significantly (by approximately two orders of magnitude on average) relative to the ground-state coupling $J_0$ between the radicals without optical excitations. This large ratio of $J_1$ to $J_0$ implies the possibility of 'switching on' a two-qubit interaction by optical excitation, thus facilitating the construction of molecular networks for quantum computing. The time evolution of the reduced density matrix for the radical suggests we could create the entanglement between the radicals through the optical excitation and ISC. We have also calculated the TREPR spectra, which is in agreement with the previous experiments. Our methodology presented here demonstrates a universal route to design QC networks by integrating nanophotonic devices and suitable molecules that can be identified through machine learning screening  for chemical database \cite{schleder2019, butler2018}, providing target molecules for chemical synthesis. The related parameters such as light intensities, excitation wavelengths, and induced exchange interactions would be the key to facilitate the choice of suitable molecules for such QC design. The blueprint proposed here will not only scale up the QC networks by orders of magnitude, but also have a great potential to operate at much higher temperature.

\section*{Data Availability}
All the computer code and data that support the findings of this study are available from the corresponding author upon reasonable request.


\begin{acknowledgments}
The authors gratefully acknowledge Prof. Hao Gong (National University of Singapore), Prof. Jing Liu (Technical Institute of Physics and Chemistry, CAS \& Tsinghua University), Prof. Gabriel Aeppli, Prof. Sandrine Heutz, Prof. Chris Kay, Dr. Garvin Morley, the late Prof. Marshall Stoneham, and Dr. Marc Warner, for helpful discussions. The authors thank the National Natural Science Foundation of China (Nos. 62164006 and 11564023), Yunnan Fundamental Research Projects (No. 202101AS070036), Yunnan Local University Joint Special Funds for Basic Research (Nos. 2017FH001-007 and 2018FH001-017), Scientific Research Fund of Yunnan Education Department (Nos. 2020Y0464, 2021Y704, and 2021Y713), UK Research Councils Basic Technology Program (EP/F041349/1), and the EU Horizon 2020 Project Marketplace (No. 760173) for funding.
\newline
\end{acknowledgments}

\section*{Author Contributions}
W. W., T. Y. Z, and H. W. contributed to the concept of the paper. J. W. C., T. H. H., L. M., and W. W. performed the theoretical analysis. All the authors contributed to the drafting of the paper. 
\newline
\section*{Competing Interests}
The authors declare no competing interests.


\begin{thebibliography}{99}
\bibitem{nielsenchuang} M. A. Nielsen, et al. \emph{Quantum computation and quantum information} (Cambridge Univ. Press, 2000).

\bibitem{alexeev2021} Y. Alexeev, et al. Quantum computer systems for scientific discovery. \emph{PRX Quantum} \textbf{2}, 017001 (2021).

\bibitem{thew2020} R. Thew, et al. Focus on quantum science and technology initiatives around the world. \emph{Quantum Sci. Technol.} \textbf{5}, 010201 (2020).

\bibitem{arute2019} F. Arute, et al. Quantum supremacy using a programmable superconducting processor. \emph{Nature} \textbf{574}, 505-510 (2019).

\bibitem{wu2021} Y. Wu, et al. Strong quantum computational advantage using a superconducting quantum processor. \emph{Phys. Rev. Lett.} \textbf{127}, 180501 (2021).

\bibitem{ebadi2021} S. Ebadi, et al. Quantum phases of matter on a 256-atom programmable quantum simulator. \emph{Nature} \textbf{595}, 227-232 (2021).

\bibitem{huang2019} W. Huang, et al. Fidelity benchmarks for two-qubit gates in silicon. \emph{Nature} \textbf{569}, 532-536 (2019).

\bibitem{xue2021} X. Xue, et al. CMOS-based cryogenic control of silicon quantum circuits. \emph{Nature} \textbf{593}, 205-210 (2021).

\bibitem{auf2022} A. Auff\'eves. Quantum technologies need a quantum energy initiative. \emph{PRX Quantum} \textbf{3}, 020101 (2022).

\bibitem{ma2022} L. Ma, et al. Triplet-radical spin entanglement: potential of molecular materials for high-temperature quantum information processing. \emph{NPG Asia Mater.} \textbf{14}, 45-54 (2022).

\bibitem{wu2023} Wei Wu, A quantum circuit architecture based on the integration of nanophotonic devices and two-dimensional molecular network, \emph{Proc. SPIE} \textbf{12335}, Quantum Technology: Driving Commercialisation of an Enabling Science III, 123350L (11 January 2023).

\bibitem{delaney2022} R. D. Delaney, et al. Superconducting-qubit readout via low-backaction electro-optic transduction. \emph{Nature} \textbf{606}, 489-493 (2022).

\bibitem{bayliss2020} S. L. Bayliss, et al. Optically addressable molecular spins for quantum information processing. \emph{Science} \textbf{370}, 1309-1312 (2020). 

\bibitem{wasielewski2020} M. R. Wasielewski, et al. Exploiting chemistry and molecular systems for quantum information science. \emph{Nat. Rev. Chem.} \textbf{4}, 490-504 (2020).

\bibitem{rv2012} I. Ratera, et al. Playing with organic radicals as building blocks for functional molecular materials, \emph{Chem. Soc. Rev.} \textbf{41}, 303 (2012).

\bibitem{ji2020} L. Ji, et al. Air-Stable Organic Radicals: New-Generation Materials  for Flexible Electronics?, \emph{Adv. Mater.} \textbf{32}, 1908015 (2020).

\bibitem{warner2013} M. Warner, et al. Potential for spin-based information processing in a thin-film molecular semiconductor. \emph{Nature} \textbf{503}, 504-508 (2013).

\bibitem{ga2019} A. Gaita-Ari$\mathrm{\tilde{n}}$o, et al. Molecular spins for quantum computation. \emph{Nat. Commun.} \textbf{11}, 301-309 (2019).

\bibitem{atzori2019} M. Atzori, et al. The second quantum revolution: role and challenges of molecular chemistry. \emph{J. Am. Chem. Soc.} \textbf{141}, 11339-11352 (2019).

\bibitem{Kjaergaard2020} M. Kjaergaard, et al. Superconducting qubits: Current state of play. \emph{Annu. Rev. Condens. Matter Phys.} \textbf{11}, 369-395 (2020).



\bibitem{atzori2016} M. Atzori, et al. Room-temperature quantum coherence and rabi oscillations in vanadyl phthalocyanine: toward multifunctional molecular spin qubits. \emph{J. Am. Chem. Soc.} \textbf{138}, 2154-2157 (2016).



\bibitem{porphyrinbooks} D. Dolphin. \emph{The porphyrins} (Academic Press, 1979).

\bibitem{teki2000} Y. Teki, et al. Intramolecular spin alignment utilizing the excited molecular field between the triplet (S =1) excited state and the dangling stable radicals (S = 1/2) as studied by time-resolved electron spin resonance: observation of the excited quartet (S = 3/2) and quintet (S = 2) states on the purely rrganic $\pi$-conjugated spin systems. \emph{J. Am. Chem. Soc.} \textbf{122} 984-985 (2000).

\bibitem{kolb2003} H. C. Kolb, et al. “The growing impact of click chemistry on drug discovery,” \emph{Drug Discovery Today} \textbf{8}, 1128–1137 (2003).

\bibitem{sato2007} O. Sato, et al. Control of magnetic properties through external stimuli. \emph{Angew. Chem. Int. Ed.} \textbf{46}, 2152-2187 (2007).

\bibitem{zhang2022} X. Zhang, et al. Electron spin resonance of single iron phthalocyanine molecules and role of their non-localized spins in magnetic interactions. \emph{Nat. Chem.} \textbf{14}, 59-65 (2022).

\bibitem{ishii1998} K. Ishii, et al. Phthalocyanine-based fluorescence probes for detecting ascorbic acid: phthalocyaninatosilicon covalently linked to TEMPO radicals. \emph{ChemComm} \textbf{47} 4932-4934 (2011).

\bibitem{corvaja2000} C. Corvaja, et al. CIDEP of fullerene C$_60$ biradical bisadducts by intramolecular triplet-triplet quenching: a novel spin polarization mechanism for biradicals. \emph{Chem. Phys. Lett.} \textbf{330}, 287-292 (2000).





\bibitem{franco2006} L. Franco, et al. TR-EPR of single and double spin labelled C$_60$ derivatives: observation of quartet and quintet excited states in solution. \emph{Mole. Phys.} \textbf{104} 1543-1550 (2006).

\bibitem{huai2005} P. Huai, et al. Electronic control of spin alignment in $\pi$-conjugated molecular magnets. \emph{Phys. Rev. B} \textbf{72},  094413 (2005).

\bibitem{bp2002} H-P. Breuer, et al. \emph{The theory of open quantum systems} (Oxford Univ. Press on Demand, 2002).




\bibitem{gaussian09} M. J. Frisch, et al. Gaussian 03  (Gaussian, Inc., Pittsburgh, PA, 1998).

\bibitem{wu2013} W. Wu, et al. Electronic structure and exchange interactions in cobalt-phthalocyanine chains. \emph{Phys. Rev. B} \textbf{88}, 024426 (2013).

\bibitem{b3lyp} A. D. Becke. Density-functional thermochemistry. III. The role of exact exchange. \emph{J. Chem. Phys.} \textbf{98}, 5648-5652 (1993).

\bibitem{wu2015} W. Wu. Hybrid-exchange density-functional theory study of the electronic structure of MnV$_2$O$_4$: Exotic orbital ordering in the cubic structure. \emph{Phys. Rev. B} \textbf{91}, 195108 (2015).

\bibitem{zou2018} T. Zou, et al. Crystal structure tuning in organic nanomaterials for fast response and high sensitivity visible-NIR photo-detector. \emph{J. Mater. Chem. C} \textbf{6}, 1495-1503 (2018).

\bibitem{ivady2020} V. Iv\'ady. Longitudinal spin relaxation model applied to point-defect qubit systems. \emph{Phys. Rev. B} \textbf{101}, 155203 (2020).

\bibitem{blankbook} A. Blank, et al. Triplet line shape simulation in continuous wave electron paramagnetic resonance experiments. \emph{Concepts Magn. Reson. A} \textbf{25}, 18-39 (2005).

\bibitem{misrabook} S. K. Misra, et al. Multifrequency electron paramagnetic resonance: theory and applications (John Wiley \& Sons, 2011).

\bibitem{easyspin} S. Stoll, et al. EasySpin, a comprehensive software package for spectral simulation and analysis in EPR, \emph{J. Magn. Reson.} \textbf{178}(1), 42-55 (2006). 

\bibitem{dpaangle} H. Noda, et al. Critical role of intermediate electronic states for spin-flip processes in charge-transfer-type organic molecules with multiple donors and acceptors. \emph{Nat. Mater.} \textbf{18}, 1084–1090 (2019). 

\bibitem{ovchinnikov1978} A. A. Ovchinnikov. Multiplicity of the ground state of large alternant organic molecules with conjugated bonds. \emph{Theor. Chim. Acta} \textbf{47}, 297-304 (1978).

\bibitem{lieb1989} E. H. Lieb. Two theorems on the Hubbard model. \emph{Phys. Rev. Lett.} \textbf{62}, 1201-1204 (1989).

\bibitem{anderson} P. W. Anderson. New approach to the theory of superexchange interactions. \emph{Phys. Rev.} \textbf{115}, 2-13 (1959).

\bibitem{ns} A. T. Boothroyd. \emph{Principles of Neutron Scattering from Condensed Matter} (Oxford Univ. Press 2020).

\bibitem{gao2013} X. Gao, et al. Quantum computer aided design simulation and optimization of semiconductor quantum dots. \emph{J. App. Phys.} \textbf{114}, 164302 (2013).

\bibitem{niquet2020} Y. M. Niquet, et al. Challenges and perspectives in the modeling of spin qubits. \emph{IEEE International Electron Devices Meeting (IEDM)} \textbf{20}, 653-656 (2020).

\bibitem{kyaw2021} T. H. Kyaw, et al. Quantum computer-aided design: digital quantum simulation of quantum processors. \emph{Phys. Rev. Appl.} \textbf{16}, 044042 (2021).

\bibitem{gibertini2019} M. Gibertini, et al. Magnetic 2D materials and heterostructures. \emph{Nat. Nanotechnol.}. \textbf{14}, 408-419 (2019).

\bibitem{ferrari2015} A. C. Ferrari, et al. Science and technology roadmap for graphene, related two-dimensional crystals, and hybrid systems. \emph{Nanoscale} \textbf{7}, 4598-4810 (2015).

\bibitem{zhuang2015} X. Zhuang, et al. Two-dimensional soft nanomaterials: a fascinating world of materials. \emph{Adv. Mater.} \textbf{27}, 403–427 (2015).

\bibitem{dsp2021} T. \DH or\dj evi\'c, et al. Entanglement transport and a nanophotonic interface for atoms in optical tweezers. \emph{Science} \textbf{373}, 1511–1514 (2021).

\bibitem{bogaerts2020} W. Bogaerts, et al. Programmable photonic circuits. \emph{Nature} \textbf{586}, 207 (2020).

\bibitem{elshaari2020} A. W. Elshaari, et al. Hybrid integrated quantum photonic circuits. \emph{Nat. Photon.} \textbf{14}, 285 (2020).

\bibitem{kern2016} J. Kern, et al. Nanoscale positioning of single-photon emitters in atomically thin WSe$_2$. \emph{Adv. Mater.} \textbf{28}, 7101–7105 (2016).

\bibitem{ziegler2018} J. Ziegler, et al. Single-photon emitters in boron nitride nanococoons. \emph{Nano Lett.}  \textbf{18}, 2683–2688 (2018).

\bibitem{peng2020} L. Peng, et al. Creation of single-photon emitters in WSe$_2$ monolayers using nanometer-sized gold tips. \emph{Nano Lett.} \textbf{20}, 5866–5872 (2020).

\bibitem{rosenberger2019} M. R. Rosenberger, et al. Quantum calligraphy: writing single-photon emitters in a two-dimensional materials platform. \emph{ACS Nano} \textbf{13}, 904–912 (2019).

\bibitem{liu2021} S. Liu, et al. Nanoscale positioning approaches for integrating single solid-state quantum emitters with photonic nanostructures. \emph{Laser Photonics Rev.} \textbf{15}, 2100223 (2021).

\bibitem{toninelli2021} C. Toninelli, et al. Single organic molecules for photonic quantum technologies. \emph{Nat. Mater.} \textbf{20}, 1615-1628 (2021).

\bibitem{benjamin2003} S. C. Benjamin, et al. Quantum computing with an always-on heisenberg interaction. \emph{Phys. Rev. Lett.} \textbf{90}, 247901 (2003).

\bibitem{schleder2019} G. R. Schleder et al, From DFT to machine learning: recent approaches to materials science–a review, \emph{J. Phys.: Mater.}, \textbf{2}, 032001 (2019). 

\bibitem{butler2018} K. T. Butler, Machine learning for molecular and materials science, \emph{Nature}, \textbf{559}, 547 (2018).




\end{thebibliography}
\end{document}